\documentclass[iop]{emulateapj}

\slugcomment{   }
\usepackage{amsmath} 
\usepackage{graphicx}
\usepackage{txfonts}
\usepackage{epsfig}
\usepackage{times}
\usepackage{natbib}
\bibpunct{(}{)}{,}{a}{}{}

\usepackage{tabularx}
\usepackage{amssymb}
\usepackage{balance}
\usepackage{natbib}
\bibpunct{(}{)}{;}{a}{}{,} % to follow the A&A style
\usepackage{color}
\usepackage{xspace}
\usepackage[normalem]{ulem} % to strike through with \sout{}
\usepackage[table]{xcolor} % for coloring the table for the referee
\usepackage{multirow}
\usepackage{rotating}
\usepackage{bbm}
\shorttitle{The red sequence at birth in Cl~J1449+0856 at $z=2$}
  
\shortauthors{Strazzullo et al.}

\begin{document}


\title{The red sequence at birth in the galaxy cluster Cl~J1449+0856 at  \lowercase{$z=2$}}


\author{V. Strazzullo$^1$, E. Daddi$^2$, R. Gobat$^3$,
  F. Valentino$^2$, M. Pannella$^1$, M. Dickinson$^4$,
  A. Renzini$^5$,\\ G. Brammer$^6$, M. Onodera$^{7,8}$, 
  A. Finoguenov$^{9,10}$,  A. Cimatti$^{11}$, C.~M. Carollo$^{12}$, N. Arimoto$^{7,8}$}

\altaffiltext{1}{Department of Physics, Ludwig-Maximilians-Universit{\"a}t,
  Scheinerstr. 1, 81679 M{\"u}nchen, Germany -- vstrazz@usm.lmu.de}  
\altaffiltext{2}{Irfu/Service d'Astrophysique, CEA Saclay,
  Orme des Merisiers, F-91191 Gif sur Yvette, France}
\altaffiltext{3}{School of
  Physics, Korea Institute for Advanced Study, Hoegiro 85,
  Dongdaemun-gu, Seoul 130-722, Republic of Korea}
\altaffiltext{4}{National
  Optical Astronomy Observatory, 950 North Cherry A venue, Tucson, AZ
  85719, USA}
\altaffiltext{5}{INAF-Osservatorio Astronomico di Padova,
  Vicolo dell'Osservatorio 5, I-35122, Padova, Italy}
\altaffiltext{6}{Space Telescope
  Science Institute, 3700 San Martin Drive, Baltimore, MD 21218, USA}
\altaffiltext{7}{Subaru Telescope, National Astronomical Observatory of Japan, National
Institutes of Natural Sciences (NINS), 650 North A'ohoku Place, Hilo, HI, 96720, USA}
\altaffiltext{8}{Department of Astronomical Science, SOKENDAI (The Graduate University for
Advanced Studies), 650 North A'ohoku Place, Hilo, HI 96720, USA}
\altaffiltext{9}{Max-Planck-Institut f\"ur extraterrestrische Physik, Giessenbachstrasse 1, 85748 Garching, Germany}
\altaffiltext{10}{Department
  of Physics, University of Helsinki, P.~O. Box 64, FI-00014,
  Helsinki, Finland}
\altaffiltext{11}{Dipartimento di Fisica e Astronomia, Universit{\'a} di Bologna,
  Viale Berti Pichat 6/2, I-30127, Bologna, Italy}
\altaffiltext{12}{Institute for Astronomy, ETH Z{\"u}rich,
  Wolfgang-Pauli-strasse 27, 8093 Z{\"u}rich, Switzerland}

\begin{abstract}
We use {\it~HST}/WFC3 imaging to study the red population in the
IR-selected, X-ray detected, low-mass cluster Cl~J1449+0856 at z=2,
one of the few {\it bona-fide} established clusters discovered at this
redshift, and likely a typical progenitor of an average massive
cluster today. This study explores the presence and significance of an
early red sequence in the core of this structure, investigating the
nature of red sequence galaxies, highlighting environmental effects on
cluster galaxy populations at high redshift, and at the same time
underlining similarities and differences with other distant dense
environments. Our results suggest that the red population in the core
of Cl~J1449+0856 is made of a mixture of quiescent and dusty
star-forming galaxies, with a seedling of the future red sequence
already growing in the very central cluster region, and already
characterising the inner cluster core with respect to lower density
environments. On the other hand, the color-magnitude diagram of this
cluster is definitely different from that of lower-redshift $z\lesssim
1$ clusters, as well as of some rare particularly evolved massive
clusters at similar redshift, and it is suggestive of a transition
phase between active star formation and passive evolution occurring in
the proto-cluster and established lower-redshift cluster regimes.

\end{abstract}


\section{Introduction}
\label{sec:intro}
\setcounter{footnote}{0} 

In the nearby Universe and up to $z\sim1$, cluster cores host the most
massive early-type galaxies containing stars nearly as old as the
Hubble time, making for a “red sequence” in their color-magnitude
diagram (CMD) \citep[e.g.,][]{kodamaearimoto,mei2009}, often
considered as a defining signature of dense environments. Observations
of the evolution of the red galaxy population back to the
proto-cluster regime show that, although massive passive galaxies and
tight red sequences in cluster cores are already common even earlier
than $z\sim1$ \citep[e.g.,][]{lidman2008,newman2014}, the red sequence
seems to increase its scatter, depopulate at lower stellar masses
\citep[note possible dependence on cluster mass,
  e.g.][]{tanaka2007,andreon2008,hilton2009,papovich2010,rudnick2012,lemaux2012,cerulo2016}
and eventually dissolve in most distant cluster progenitor
environments. A first sequence probably appears at high stellar masses
just before $z\sim2$ \citep{kodama2007,zirm2008}, in agreement with
star formation histories inferred for lower-redshift passive cluster
galaxies
\citep[e.g.,][]{gobat2008,rettura2010,strazzullo2010,snyder2012},
placing their major star formation epoch at $z_{f}\gtrsim2$.

In the last few years, cluster studies systematically approached this
expected formation epoch and indeed started revealing increasing star
formation, nuclear and merging activity in $z\gtrsim1.5$ clusters
\citep[e.g.,][]{hilton2010,hayashi2011,stanford2012,brodwin2013,tran2015,santos2015,webb2015,alberts2016}. \citet{wang2016}
study of a $z=2.5$ structure seemingly captures the massive star
formation event likely producing the future red sequence.  On the
other hand, although often co-existing with still actively
star-forming (SF) sources, quiescent galaxies are found in cluster
cores even up to $z\sim2$
\citep[e.g.,][]{kurk2009,gobat2011,gobat2013,spitler2012,tanaka2013,strazzullo2013,smail2014,newman2014,cooke2016},
with enhanced quiescent fractions with respect to field levels
testifying to an early onset of environmental signatures on galaxy
population properties.

\begin{figure*}[ht!]
 \includegraphics[width=\textwidth]{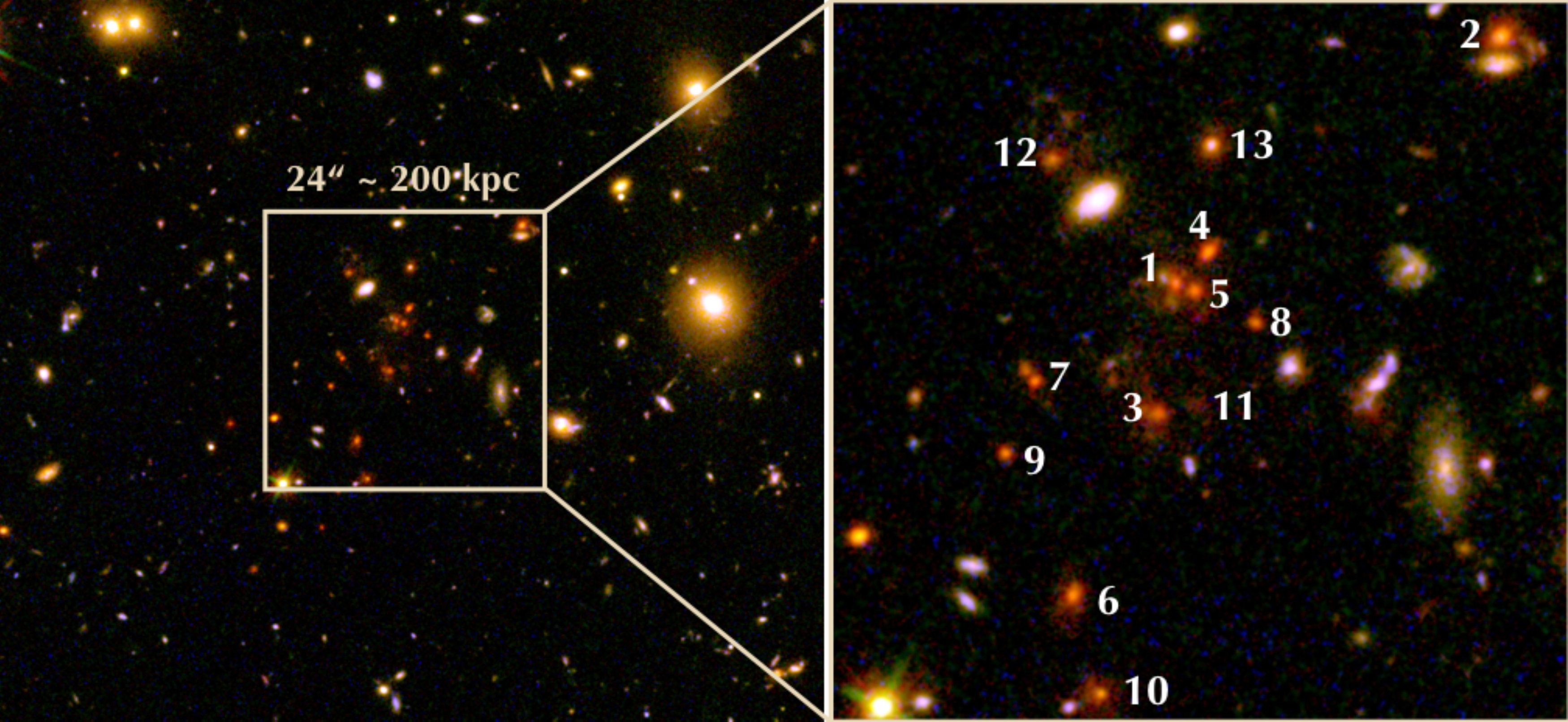}
\caption{F606W-F105W-F140W pseudo-color image of ClJ1449 (left,
  $\sim 60'' \times 70''$), and a zoom into the cluster center (right,
  $\sim200\times200$~kpc$^2$ (proper) at z=2). In the right-hand
  panel, numbers mark red sources discussed in this work.
\label{fig:colima} }
\end{figure*}

In this work, we follow-up on our previous studies of the cluster
Cl~J1449+0856 (hereafter ClJ1449) at z=1.995 \citep[][hereafter G11,
  G13, S13, V15,
  V16]{gobat2011,gobat2013,strazzullo2013,valentino2015,valentino2016}
with an investigation of the CMD based on {\it Hubble Space Telescope}
({\it~HST}) observations. These data provide a deep, high-resolution probe
of the restframe $\sim$U-B color of galaxies in ClJ1449, significantly
improving our previous photometric analyses and eventually enabling a
study of the red population in one of the most distant cluster
environments known thus far.

ClJ1449 is a low-mass ($\sim5\cdot10^{13}$~M$_{\odot}$), IRAC-color
selected, X-ray detected structure in the mass range of typical
progenitors of today's massive clusters (G11). In contrast with rare
examples of impressively evolved structures at similar redshift
\citep{andreon2014,newman2014}, ClJ1449 thus probably draws a more
common picture of galaxy populations in the first clusters at this
crucial epoch.

We assume $\Omega_{M}$=0.3,
$\Omega_{\Lambda}$=0.7, H$_{0}$=70~km~s$^{-1}$~Mpc$^{-1}$, and a
\citet{salpeter1955} IMF. Magnitudes  are in the AB system.

\section{Observations and measurements}

This work is largely based on {\it~HST} Wide Field Camera 3 (WFC3)
observations of ClJ1449 in the F105W and F140W
bands. Fig.~\ref{fig:colima} shows an image of the cluster field, and
highlights all potential red cluster members in the central region
(m105-m140$>$0.9, $r<200$~kpc). We further use optical/NIR data and
derived quantities from our previous work, and specifically
spectroscopic redshifts (G11, G13, V15) to identify cluster
members, complemented with photometric redshift (photo-z) based
membership (S13, see original papers for details). In the
following we refer to: {\it i) interlopers} including spectroscopic
and photo-z interlopers, {\it ii) possible members} including
uncertain, low-likelihood photo--z candidate members (S13), and {\it iii)
  members} including spectroscopic and high-likelihood photo-z members
(S13). We use photometric measurements and derived quantities from
S13, complemented where needed with new measurements performed with
the same approach accounting for the new F105W
data\footnote{Measurements for some galaxies will thus differ from
  those in S13. We note (in particular for source ID~1, see
  Fig.~\ref{fig:colima}) that the new F105W data support in some
  (rare) case a different source extraction than the one used in S13,
  resulting in different source properties.  Furthermore, we
  redetermined stellar masses using the newly available photometry,
  and - in contrast to S13 - using the original photometric
  zero-points rather than those adjusted for photo-z
  determination. Stellar masses used here are thus not identical to
  those in S13.}.

We also use a control field of $\sim$60~arcmin$^2$ in the CANDELS
\citep{grogin2011} GOODS-S field as a statistical comparison for
galaxy populations in ClJ1449. We use multiwavelength photometry from
\citet{guo2013} and photo--zs from \citet{schreiber2015}. Lacking
sufficiently deep F140W-band imaging, we measure synthetic m140
magnitudes by convolving best-fit SEDs from \citet{pannella2015} with
the F140W  response function. In the magnitude range of
interest, we estimate a $\lesssim0.1$~mag uncertainty for these
synthetic magnitudes, which is not relevant for our purposes.

\begin{figure*}[htp!]
\includegraphics[height=0.229\textheight,clip,bb=62 435 558 720]{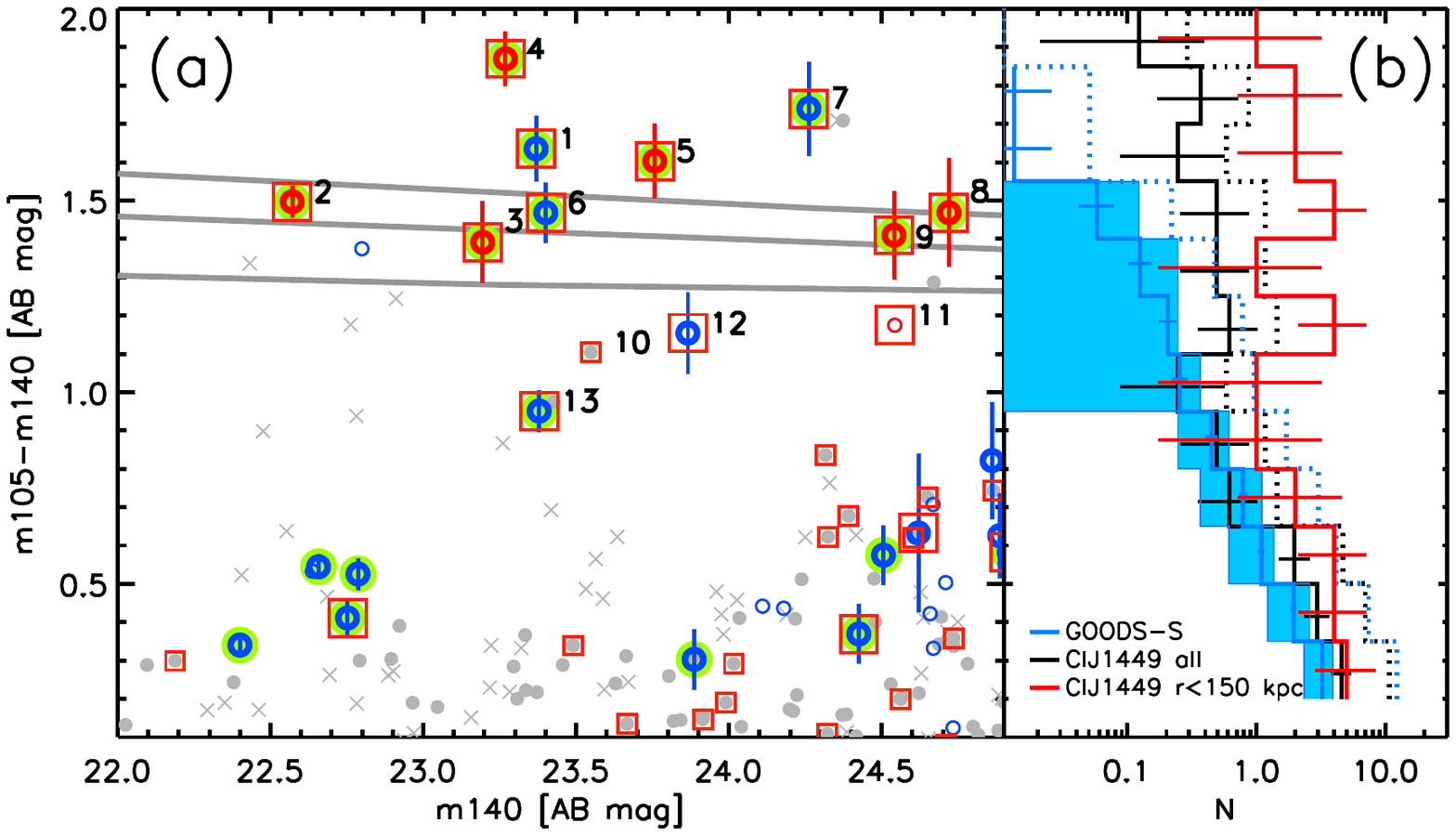}%
\includegraphics[height=0.229\textheight,clip,bb=104 435 558 720]{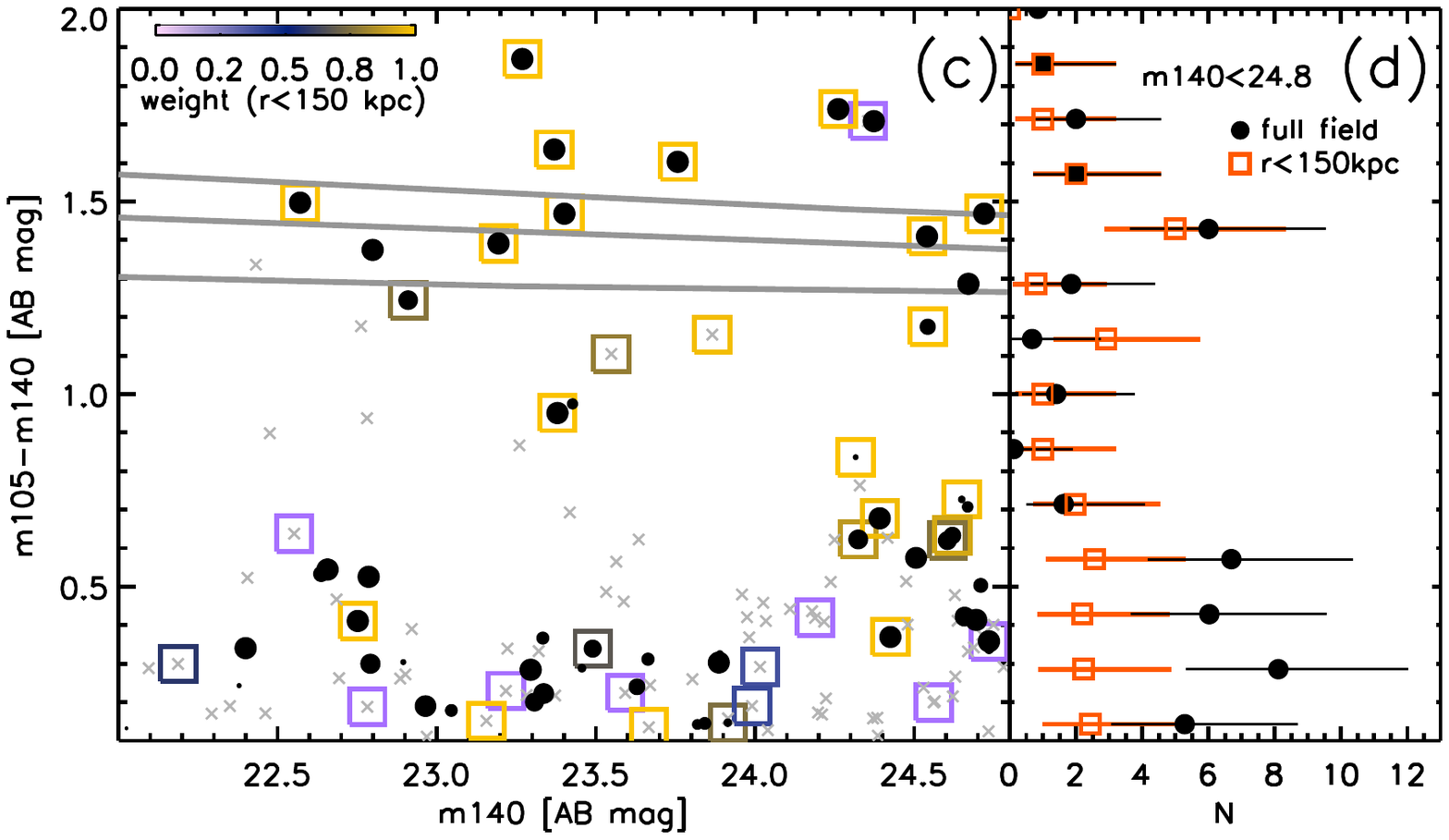}
\caption{{\bf Panel a):} CMD of all galaxies in the ClJ1449 field
  (F140W MAG\_AUTO vs. F105W-F140W aperture color, 0.6''
  diameter). Thick colored symbols show cluster members; spectroscopic
  members are highlighted with green shading. Low-likelihood candidate
  members are shown with small colored symbols, and photometric and
  spectroscopic interlopers with gray points and crosses,
  respectively. Color-coding of members reflects the classification
  from Fig.~\ref{fig:colcol} (see text). Red squares mark all galaxies
  (but spectroscopic interlopers) within 150~kpc from the cluster
  center. Red sources are numbered as in Fig.~\ref{fig:colima}.  The
  three gray lines show the KA97 model red sequence at z=2 for $z_f=
  3,5,10$.  {\bf Panel b): } The color histogram of {\it all}
  $22<m140<24.8$ galaxies in the ClJ1449 field (black, solid), within
  $r<150$~kpc from the cluster center (red), and in the GOODS-S
  control field (blue, solid), all normalised to the projected area
  within $r<150$~kpc. The cyan-shaded area around the blue line shows
an estimate of cosmic variance from cluster field-sized apertures
randomly placed over the control field. Dotted lines show histograms
normalised to the total number of galaxies within $r<150$~kpc. {\bf
  Panel c): } A visual rendition of the CMD based on statistical
background subtraction (see text). Black points show {\it all}
galaxies in the ClJ1449 field, with symbol size scaling with their
excess probability over the CMD in the control field (galaxies with
zero excess probability are plotted as gray crosses - see text for details). 
Coloured squares show this statistically
background-subtracted CMD in the central $r<150$~kpc; symbol color
scales with excess probability according to the upper-left
colorbar. {\bf Panel d):} The statistically background-subtracted
(from panel c) color distribution of cluster galaxies, in the full
ClJ1449 field and within $r<150$~kpc.
\label{fig:cmd} }
\end{figure*}

\section{The color-magnitude diagram}

Fig.~\ref{fig:cmd}a shows the F105W-F140W CMD in the field of ClJ1449,
over a region extending to a maximum clustercentric distance r=500~kpc
(proper) at $z=2$ ($\sim$1', about the estimated
$r_{200}\sim0.78^\prime\sim0.4\pm0.1$~Mpc for ClJ1449, G13, V16).
  All galaxies in the probed field are shown, but cluster members are
  highlighted with colored symbols according to their stellar
  population properties (Sec.~\ref{sec:stellarpops},
  Fig.~\ref{fig:colcol}).

Fig.~\ref{fig:cmd}c shows an alternative, photo--z independent picture
of the CMD, largely based on statistical background subtraction
(individual membership is only considered for spectroscopic sources,
and only affects Fig.~\ref{fig:cmd}c in minor details). This CMD is
obtained by subtracting, for each galaxy in the control (GOODS-S)
field, an area-normalised fractional galaxy contribution to the
closest galaxy in color-magnitude space in the ClJ1449 field
\citep[see e.g.][]{vanderburg2016b}. All galaxies in the ClJ1449
field are shown (black/gray and colored symbols for the full-field and
$r<150$~kpc CMD, respectively), but symbol size/color scales with
their excess probability over the control field CMD. The excess
probability\footnote{Being {\it statistical}, the excess probability
  in Fig.~\ref{fig:cmd}c obviously does not directly reflect cluster
  membership probability on a galaxy-by-galaxy basis.}  across the CMD
of Fig.~\ref{fig:cmd}c is meant to represent the background-subtracted
color-magnitude distribution of cluster galaxies in ClJ1449, as an
independent confirmation of the color distribution based on individual
membership shown in Fig.~\ref{fig:cmd}a.

As compared with $z\lesssim 1$ clusters, or even with the massive
$z=1.8$ cluster JKCS041 \citep{newman2014}, the CMD of ClJ1449 shows a
clearly less dominant, more scattered red galaxy
population. Nonetheless, a first albeit still sparsely populated red
sequence is found close to the expected location, confirming the
presence of a characteristic population of red sources in the cluster
core (G11, G13, S13), and further highlighting red-sequence
outliers. As a reference, the red sequence predicted at z=2 by
\citet[][KA97]{kodamaearimoto} models is also shown.  An estimate of
the intrinsic m105-m140 scatter (restframe $\sim$U-B)
\citep[determined as in][in our case affected by very poor
  statistics]{lidman2008} of {\it all} core ($r<150$~kpc) red members
with m105-m140$>$1.2 (regardless of their quiescent/SF nature, see
Sec.~\ref{sec:stellarpops}), is $\sigma_{int}=0.14^{+0.03}_{-0.07}$
($\sigma_{int}=0.12\pm0.03$ if considering only core members
consistent within 0.15~mag with the KA97 $3<z_f<10$ models).  On the
other hand, with the exception of ID~4 {\it quiescent} members lie on
a tight sequence with $z_{f}\gtrsim5$ \citep[see
  Secs.~\ref{sec:stellarpops},~\ref{sec:conclusions}, see also
  e.g.][]{stanford2012}.

Fig.~\ref{fig:cmd}b and \ref{fig:cmd}d show, respectively, the color
distribution of {\it all} $22<m140<24.8$ galaxies in the ClJ1449 field
(full field and $r<150$~kpc), and of {\it cluster} galaxies
(statistically background subtracted as from
Fig.~\ref{fig:cmd}c). Fig.~\ref{fig:cmd}b also shows for comparison
the color distribution  of {\it all} $22<m140<24.8$ galaxies in
  the control field. Figures ~\ref{fig:cmd}b and ~\ref{fig:cmd}d are
not affected by uncertainties in the individual membership
determination, as they use the full galaxy sample or a statistically
background-subtracted sample (as for Fig.~\ref{fig:cmd}c). They thus
provide a further confirmation and quantification of the excess
population of red sources in the cluster field shown in
Fig.~\ref{fig:cmd}a.

Besides the very clear excess of red galaxies, Fig.~\ref{fig:cmd}
generally suggests that the ``green valley'' is still indeed an
underdense region in the CMD of ClJ1449 over the full ($r<500$~kpc)
area probed, with the excess of red sources being located at markedly
redder colors, thus constraining the relevance of the population
transitioning from the ``blue cloud'' at bright magnitudes.   On the
other hand, although statistics are very poor, we note that virtually
all green-valley sources likely belonging to the cluster are located
within $r<150$~kpc from the cluster center.

\section{The red  population}
\label{sec:stellarpops}

\begin{figure}[t]
\includegraphics[width=0.48\textwidth,clip,bb=62 360 362 720]{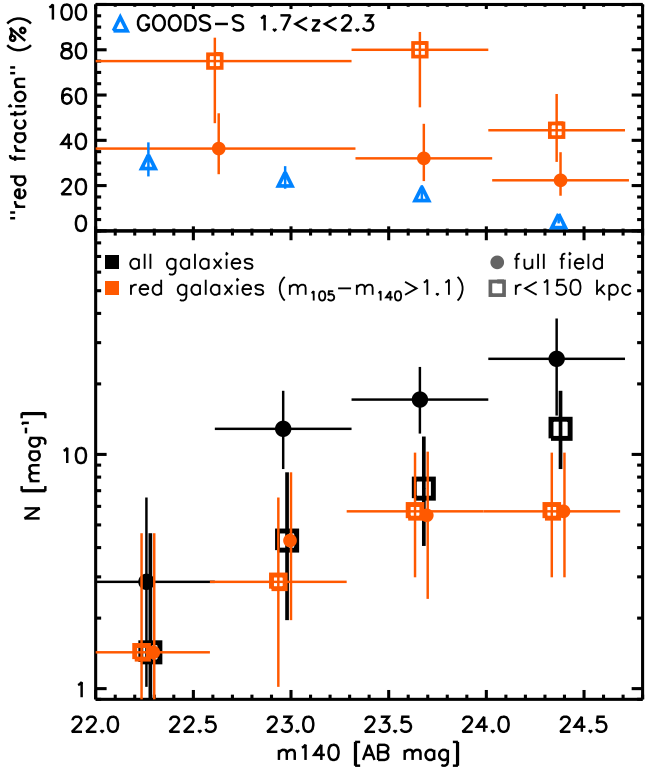}
\caption{{\bf Bottom panel: } The F140W LF in the field of
  ClJ1449. Filled circles and empty squares show the LF in the full
  field and within $r<150$ kpc. Black and red symbols refer to the
  full sample and to red galaxies, respectively (see text for
  details). Error bars show Poisson errors.  {\bf Top panel: } Red
  points show the red fraction in ClJ1449 from the ratio of the red
  and total LFs in the bottom panel.  Blue points show the
  corresponding red fraction in the GOODS-S control field for galaxies
  at $1.7<z<2.3$. Errors are calculated following
  \citet{cameron2011}.
\label{fig:LF} }
\end{figure}

In order to further highlight the relevance of the red population in
ClJ1449, Fig.~\ref{fig:LF} compares the F140W luminosity functions
(LFs) of all and red galaxies in the ClJ1449 field, and in the central
$r<150$~kpc area. The LF is essentially determined through statistical
background subtraction using the GOODS-S control field \citep[though
  we make use of spectroscopic and photometric redshift information to
  remove obvious interlopers, in particular at the bright end where
  statistics are very poor, e.g.,][]{strazzullo2010}.  The ``red''
galaxy LF is obtained selecting galaxies with m105-m140$>$1.1 and,
here again, includes the full red population regardless of stellar
population properties discussed below. Fig.~\ref{fig:LF} also shows
the corresponding red fraction as a function of m140 magnitude,
determined as the ratio of the red and the total LFs, and compared
with the analogous red fraction for $1.7<z_{phot}<2.3$ galaxies in the
GOODS-S control field.  Like Fig.~\ref{fig:cmd}, Fig.~\ref{fig:LF}
shows again that the red population is enhanced only in the very
central cluster region, the red fraction in the overall cluster virial
volume being similar to the control field levels.

\begin{figure}[b]
\includegraphics[width=0.48\textwidth,clip,bb=62 436 413 720]{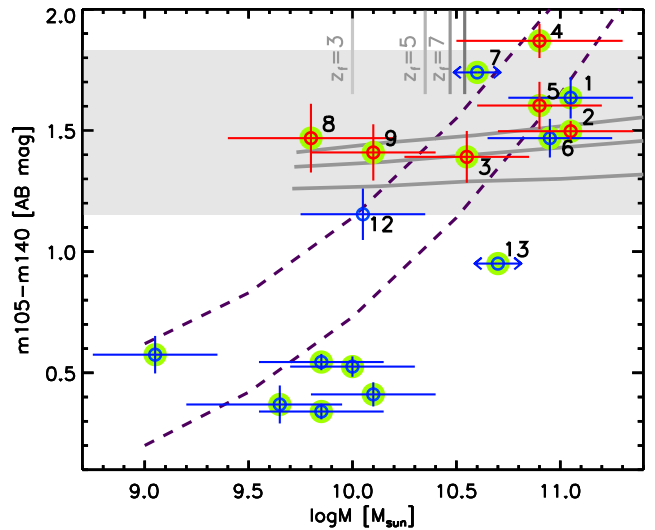}
\caption{Color-mass diagram of cluster members in the magnitude range
  $22<$~m140~$<24.7$ (see Fig.~\ref{fig:cmd}). Color coding follows
  Fig.~\ref{fig:cmd}. Vertical gray lines show the stellar mass of a
  m140=24.7, solar-metallicity BC03 SSP  with different formation
  redshifts: the darkest line corresponds to a maximally old
  population, approximately marking the mass completeness of the
  sample for passive dust-free galaxies. The gray area shows the
  color range of passive BC03 SSPs from Fig.~\ref{fig:colcol}. Gray
  slanted lines show KA97 models from Fig.~\ref{fig:cmd}. Dashed
  purple lines show, as a reference, constant-SFR BC03 models of
  age=0.1, 1 Gyr, affected by mass-dependent dust reddening as from
  \citet{pannella2015}.
\label{fig:colmass} }
\end{figure}

Finally, Fig.~\ref{fig:colmass} shows the color-stellar mass diagram
for cluster members, highlighting that - as expected - optically-blue
galaxies are typically restricted to low stellar masses, while the
high-mass population is typically dominated by red sources (old or
dusty, see Sec.~\ref{sec:stellarpops} below).

The actual nature of the red population shown in Fig.~\ref{fig:cmd}
deserves further discussion. Sources labeled as IDs 1, 4 and 5 in
Fig.~\ref{fig:cmd}a are part of what G11 identified as a ``proto-BCG''
still in active formation (see Fig.~\ref{fig:colima}, right). The
high-resolution deep {\it~HST} imaging, as well as ALMA observations
(Strazzullo et al., in prep.), indeed reveal a complex structure with
multiple components. According to photometric and spectroscopic
analysis, S13 and G13 classified IDs 1 and 4 of this proto-BCG complex
as (dusty) SF and quiescent, respectively; IDs 2, 3, 6, 8, and 9 were
also classified as quiescent.  The sample of ``red sources''
highlighted in Fig.~\ref{fig:colima} further includes four
green-valley sources, IDs~10 to 13; based on photo--zs, IDs~10 and 11
are probably interlopers.

 We note that IDs~7 and 13 are AGN hosts (G13) which may affect their
 colors and derived properties (together with significant neighbour
 contamination of ground-based and IRAC photometry for ID~7). Although
 we include them for completeness in the figures, a proper analysis of
 the nature of these two sources is postponed to future work and we
 thus do not consider them further in the following discussion.

\begin{figure}[t]
\includegraphics[width=0.45\textwidth,clip,bb=63 435 360 720]{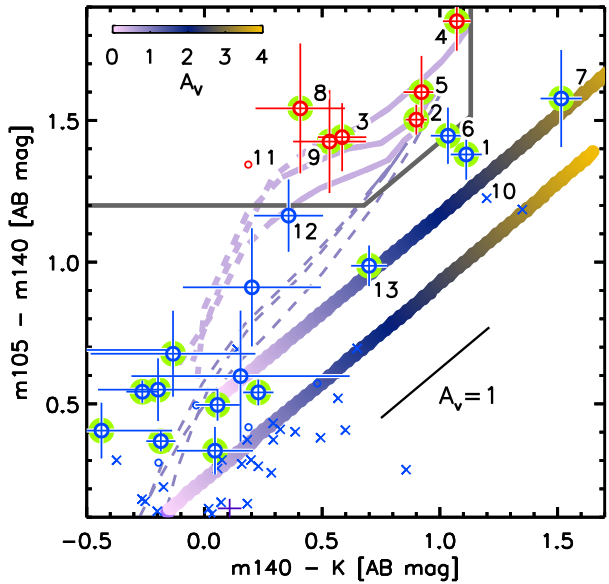}
\caption{The m105-m140 vs. m140-K plot for  cluster
  members (symbol coding as in Fig.~\ref{fig:cmd}; photo--z interlopers
  are still shown as crosses). The thick colored lines show for
  reference the colors of two BC03 constant-SFR models at $z=2$, with
  ages 0.1 and 1Gyr, and dust attenuation according to the top-left
  color scale. Thin purple lines show the age evolution of models with
  $\tau = 0$ and 0.5 Gyr, $0.008<Z<0.05$ and A$_{V}$=0; lines are
  dashed until age$>$0.5 Gyr and age/$\tau>$4, then become solid lines
  (see text). The gray line defines the adopted region of ``passive
  galaxies'' (upper left part of the plot), determining the
  color-coding of points: red/blue for galaxies in the
  passive/star-forming regions, respectively.
\label{fig:colcol} }
\end{figure}

The new F105W imaging allows a further photometric classification of
galaxy stellar populations, similar in principle to the UVJ or p/sBzK
approaches \citep{williams2009,daddi2004} aiming at a separation of
quiescent and SF galaxies accounting for dust reddening, and
independent of SED modeling (as we used in S13) which may be
particularly affected by blending in the crowded cluster center when
using imaging with poorer resolution. Fig.~\ref{fig:colcol} shows the
m140-K vs m105-m140 color-color plot for cluster members (1'' diameter
aperture colors)\footnote{Because of the combination with ground-based
  K-band imaging, this aperture is larger than the one used in
  Fig.~\ref{fig:cmd}. Although color gradients between these two
  apertures are generally small, there are few exceptions, notably
  IDs~1 and 7; given the complex surroundings of these sources this
  might be partly due to contamination by other sources/components.}.

We consider the position of cluster galaxies in this diagram to
constrain the nature of their stellar populations.  As an indicative
reference of the location of SF populations in this diagram, we show
the colors of two constant star formation rate (SFR) BC03 models at
$z=2$ with ages 0.1 and 1~Gyr, attenuated according to
\citet{calzetti2000} with $0<A_{V}<4$.  The evolution of
(dust free) models with $\tau=0$ (SSP) and 0.5 Gyr, and
$0.008<Z<0.05$, is also shown, for ages up to the age of the Universe
at $z=2$.  As Fig.~\ref{fig:colcol} shows, this {\it observed} color
combination still allows for the definition of a quiescent region with
a boundary parallel to the reddening vector, enabling a classification
of stellar populations only relying on observed magnitudes from
imaging with relatively good resolution.

For the vast majority of sources, the quiescent/SF classification from
Fig.~\ref{fig:colcol} agrees with that UVJ/SED-based from S13. The
notable exception relevant for this work is ID~6 (and partly ID13 with
an uncertain classification in S13). In Fig.~\ref{fig:colcol}, IDs~1,
6 and 12 are formally in the SF region, although very close to the
boundary line, thus their m105-m140-K classification may be deemed
uncertain.  However, the recently identified ALMA continuum and
CO(4-3) emission from ID~6 supports its dusty SF nature.  Concerning
ID~1, as discussed in G11 {\it Spitzer}/MIPS 24$\mu$m emission was
observed close to the location of the proto-BCG. However, the
ALMA-detected 870$\mu$m continuum and CO(4-3) emission is clearly
associated with a very red faint component south of the main bright
sources discussed here, IDs~1, 4, 5, which are instead all below the
870$\mu$m map $2\sigma$ level \citep[0.2mJy, $\sim$40~M$_{\odot}$/yr,
  e.g.][]{bethermin2012}, as all other red and green valley sources
discussed here besides ID~6. Given the estimated stellar masses and
colors, this may suggest that most of the massive red sources IDs~1,
2, 3, 4, 5 are likely quiescent/suppressed-SF \citep[at least with
  respect to the expected mass-SFR relation, e.g.][]{elbaz2011} rather
than dusty massively SF galaxies. We recall nonetheless the
significant uncertainties inherent to this kind of analysis; a more
detailed discussion of the ALMA observations will be presented in
Strazzullo et al., in preparation. For IDs~8, 9, 12, 13, given the
expected stellar masses and SFRs, the available ALMA observations do
not add significant constraints.

\section{Conclusions}
\label{sec:conclusions}

We use {\it~HST}/WFC3 observations to further investigate the population of
optically-red galaxies in the cluster ClJ1449 at z=2. This work
highlights once more the presence of a numerically small but
characteristic population of red galaxies concentrated within
$r<150$~kpc from the cluster center.
Many of these red sources lie along a ``red sequence'' which is close
to passive evolution expectations at this redshift, however this
sequence is significantly more scattered than in $z\lesssim1$ clusters
\citep[e.g.][]{mei2009,foltz2015}, notably with the presence of
extremely red sources, and is partly contaminated by dusty SF
galaxies. The red population in ClJ1449 is more sparse and scattered
even as compared with some (rather exceptional) cluster at similar
redshift \citep[notably JKCS041 at $z=1.8$,][]{newman2014}, although we
also note results from
e.g.~\citet{hilton2010,papovich2010,stanford2012,cerulo2016} and references
therein on the marked increase of cluster red-sequence scatter at
$z\gtrsim1.4$.

Even among the red galaxies that are deemed to be quiescent rather
than dusty SF, two sources scatter above the KA07 $z_f\sim10$
red-sequence prediction (IDs~4 and 5), suggesting dust attenuation
and/or supersolar metallicity (see Fig.~\ref{fig:colcol}), although we
recall the especially crowded environment of these specific galaxies
possibly affecting photometric accuracy. As a reference, the colors of
ID~4 may be reproduced with a young (age$\sim$0.5 to $\sim$1~Gyr)
solar-metallicity BC03 SSP with $A_{V}\sim1.5-1$, or possibly with a
dust-free super-solar (Z=0.05) SSP with a $z_f\gtrsim6$.

Early and recent results on the investigation of optically red
galaxies in dense environments at $z\gtrsim2$ highlighted the -
sometimes dominant - contribution of dusty, highly SF sources
\citep[e.g.][]{zirm2008,wang2016}.
The analysis presented here, also with the support of recent ALMA
observations, suggests that indeed ClJ1449 hosts in its very central
region a mixture of quiescent and dusty SF galaxies, at the same time
highlighting a population that is seemingly dominated in number by
galaxies with suppressed star formation making for a first, still
forming red-sequence seedling.  In combination with previous work,
this study underlines differences and similarities in galaxy
population properties of cluster progenitors of different masses in a
crucial redshift range, providing the picture of one of the most
distant spectroscopically confirmed {\it bona-fide} cluster
environments with an estimated mass placing it among the average
progenitors of today's clusters rather than exceptionally massive
structures.

~\\

% -----------------------------------------------------------------------------
\acknowledgments

  VS, ED, RG and FV were supported by grants ERC-StG~UPGAL~240039 and
  ANR-08-JCJC-0008. 
Based on observations from {\it~HST} programs GO-11648 and
GO-12991.

% -----------------------------------------------------------------------------

%\bibliographystyle{aa}
%\bibliography{ms}

\end{document}